\documentclass[sigconf]{acmart}
\usepackage{graphicx}
\usepackage{caption}
\usepackage{subcaption}
\usepackage{amsmath}
\usepackage{amssymb}
\usepackage{amsthm}
\usepackage{color}
\usepackage{ifpdf}
\usepackage{booktabs} 
\usepackage{listings}
\setcopyright{rightsretained}

\newcommand{\up}{\vspace*{-1em}}
\newcommand{\upe}{\vspace*{-0.8em}}
\newcommand{\uph}{\vspace*{-0.6em}}
\newcommand{\upp}{\vspace*{-0.4em}}

\usepackage{hyperref}
\usepackage{float}
\usepackage[utf8]{inputenc}
\usepackage{multirow}
\usepackage{rotating}
\usepackage{setspace}
\usepackage{paralist}
\usepackage{makecell}
\usepackage{moresize}
\usepackage{tabularx}
\usepackage{algorithm}
\usepackage{algpseudocode}

\usepackage{url}
\usepackage{booktabs}
\usepackage{listings}
\usepackage{paralist}
\usepackage{wrapfig}
\usepackage{multirow}
\usepackage{ifpdf}
\usepackage{xspace}
\usepackage{keyval}
\usepackage{color}
\usepackage{nicefrac}
\usepackage{float}
\usepackage{draftwatermark}

\theoremstyle{definition}

\definecolor{listinggray}{gray}{0.95}
\definecolor{darkgray}{gray}{0.7}
\definecolor{commentgreen}{rgb}{0, 0.4, 0}
\definecolor{darkblue}{rgb}{0, 0, 0.4}
\definecolor{middleblue}{rgb}{0, 0, 0.7}
\definecolor{darkred}{rgb}{0.4, 0, 0}
\definecolor{brown}{rgb}{0.5, 0.5, 0}

\usepackage[normalem]{ulem}
\def\cyanuwave{\bgroup \markoverwith{\lower3.5\p@\hbox{\sixly \textcolor{cyan}{\char58}}}\ULon}
\def\reduwave{\bgroup \markoverwith{\lower3.5\p@\hbox{\sixly \textcolor{red}{\char58}}}\ULon}
\def\blueuwave{\bgroup \markoverwith{\lower3.5\p@\hbox{\sixly \textcolor{blue}{\char58}}}\ULon}
\font\sixly=lasy6 
\makeatother

\acmDOI{ }

\acmISBN{ }

\acmConference[ICPP]{47th International Conference on Parallel Processing, August 13-16, 2018, University of Oregon, Eugene, Oregon, USA}
\acmYear{2018}
\copyrightyear{2018}

\newif\ifdraft
\ifdraft
\usepackage{xcolor}
\definecolor{ocolor}{rgb}{1,0,0.4}
\newcommand{\terminology}[1]{ {\textcolor{red} {(Terminology used: \textbf{#1}) }}}

\newcommand{\jhanote}[1]{ {\textcolor{red} { ***shantenu: #1 }}}
\newcommand{\alnote}[1]{ {\textcolor{blue} { ***andreL: #1 }}}
\newcommand{\todo}[1]{ {\textcolor{brown} { TODO #1 }}}
\newcommand{\obnote}[1]{ {\textcolor{cyan} { ***oliver: #1 }}}
\newcommand{\rthreenote}[1]{ {\textcolor{cyan} { Reviewer 3: #1 }}}
\definecolor{orange}{rgb}{1,.5,0}
\definecolor{dandelion}{cmyk}{0,0.29,0.84,0}
\newcommand{\mknote}[1]{ {\textcolor{dandelion} { ***mahzad: #1 }}}
\newcommand{\rtwonote}[1]{ {\textcolor{dandelion} { ***Reviewer 2: #1 }}}
\newcommand{\gpnote}[1]{{\textcolor{green} {***giannis: #1}}}
\newcommand{\note}[1]{ {\textcolor{magenta} { ***Note: #1 }}}
\newcommand{\rfournote}[1]{ {\textcolor{orange} { ***Reviewer 4: #1 }}}
\else
\newcommand{\terminology}[1]{}

\newcommand{\alnote}[1]{}
\newcommand{\mknote}[1]{}
\newcommand{\obnote}[1]{}
\newcommand{\rthreenote}[1]{}
\newcommand{\rfournote}[1]{}
\newcommand{\todo}[1]{}
\newcommand{\jhanote}[1]{}
\newcommand{\rtwonote}[1]{}
\newcommand{\gpnote}[1]{}
\newcommand{\note}[1]{}
\fi

\newcommand{\rp}{RADICAL-Pilot\xspace}

\newcommand{\computeunit}{Compute-Unit\xspace}
\newcommand{\computeunits}{Compute-Units\xspace}

\newcommand{\cu}{CU\xspace}
\newcommand{\cus}{CUs\xspace}

\newcommand{\showDOI}[1]{\unskip}
\newcommand{\showURL}[1]{\unskip}

\floatstyle{ruled}
\newfloat{scheme}{htbp}{lok}
\floatname{scheme}{Scheme}

\lstdefinestyle{myListing}{
    frame=single,
    backgroundcolor=\color{listinggray},
    language=C,
    basicstyle=\ttfamily \footnotesize,
    breakautoindent=true,
    breaklines=true
    tabsize=2,
    captionpos=b,
    aboveskip=0em,
    belowskip=-2em,
}

\lstdefinestyle{myPythonListing}{
    frame=single,
    backgroundcolor=\color{listinggray},
    language=Python,
    basicstyle=\ttfamily \scriptsize,
    breakautoindent=true,
    breaklines=true
    tabsize=2,
    captionpos=b,
}

\ifpdf
\DeclareGraphicsExtensions{.pdf, .jpg, .tif}
\else
\DeclareGraphicsExtensions{.ps,  .eps, .jpg}
\fi

\usepackage{listings}
\usepackage{paralist}
\lstnewenvironment{code}[1][]%
{
    \noindent
    \minipage{1.0 \linewidth}
    \vspace{0.5\baselineskip}
    \lstset{
        language=Python,
        frame=single,
        captionpos=b,
        stringstyle=\ttfamily,
        basicstyle=\scriptsize\ttfamily,
        showstringspaces=false,#1}
}
{\endminipage}

\defaultleftmargin{1em}{}{}{}

\begin{document}
    
\title{Task-parallel Analysis of Molecular Dynamics Trajectories}
    
\author{Ioannis Paraskevakos}
\affiliation{%
    \institution{Rutgers University}
    \city{Piscataway}
    \state{New Jersey}
    \postcode{08904}
    \country{USA}
}

\author{Andre Luckow}
\affiliation{%
    \institution{Ludwig-Maximilians-University}
    \city{Munich}
    \country{Germany}
}
\author{Mahzad Khoshlessan}
\affiliation{%
    \institution{Arizona State University}
    \city{Tempe}
    \state{Arizona}
    \postcode{85281}
    \country{USA}
}
\author{George Chantzialexiou}
\affiliation{%
    \institution{Rutgers University}
    \state{New Jersey}
    \postcode{08904}
    \country{USA}
}

\author{Thomas E. Cheatham}
\affiliation{%
    \institution{University of Utah}
    \city{Salt Lake City}
    \state{Utah}
    \postcode{84112}
    \country{USA}
}

\author{Oliver Beckstein}
\affiliation{%
    \institution{Arizona State University}
    \city{Tempe}
    \state{Arizona}
    \postcode{85281}
    \country{USA}
}

\author{Geoffrey C. Fox}
\affiliation{%
    \institution{Indiana University}
    \city{Bloomington}
    \state{Indiana}
    \postcode{470405}
    \country{USA}
}

\author{Shantenu Jha}
\affiliation{%
    \institution{Rutgers University and Brookhaven National Laboratory.}
     }


\renewcommand{\shortauthors}{I. Paraskevakos et al.}
    
\begin{abstract} 
	Different parallel frameworks for implementing data analysis applications
	have been proposed by the HPC and Big Data communities. 
	In this paper, we investigate three task-parallel
	frameworks: Spark, Dask and \rp with respect to their ability
	to support data analytics on HPC resources and compare them to MPI.
	We investigate the data analysis requirements of Molecular Dynamics (MD) 
	simulations which are significant consumers of supercomputing cycles, producing immense
	amounts of data. A typical large-scale MD simulation of a physical system
	of $O(100k)$ atoms over $\mu secs$ can produce from $O(10)\:GB$ to $O(1000)\:GBs$ of
	data. We propose and evaluate different
	approaches for parallelization of a representative set of MD trajectory
	analysis algorithms, in particular the computation of path similarity and
	leaflet identification. We evaluate Spark, Dask and \rp with
	respect to their abstractions 
	and runtime engine capabilities to support these
	algorithms. We provide a conceptual basis for comparing and understanding
	different frameworks that enable users to select the optimal system for
	each application. We also provide a quantitative performance 
	analysis of the different algorithms across the three frameworks.
\end{abstract}


\copyrightyear{2018} 
\acmYear{2018} 
\setcopyright{acmcopyright}
\acmConference[ICPP 2018]{47th International Conference on Parallel Processing}{August 13--16, 2018}{Eugene, OR, USA}
\acmBooktitle{ICPP 2018: 47th International Conference on Parallel Processing, August 13--16, 2018, Eugene, OR, USA}
\acmPrice{15.00}
\acmDOI{10.1145/3225058.3225128}
\acmISBN{978-1-4503-6510-9/18/08}

\maketitle
\section{Introduction}
Frameworks for parallel data analysis have been created by the High Performance 
Computing (HPC) and Big Data communities~\cite{fox-2017}. MPI is the most used programming 
model for HPC resources. It assumes a SPMD execution model where each process 
executes the same program. It is highly optimized for high-performance computing 
and communication, which along with synchronization need explicit implementation. 
Big Data frameworks utilize higher-level MapReduce~\cite{mapreduce} programming 
models avoiding explicit implementations of communication. In addition, the 
MapReduce~\cite{mapreduce} abstraction makes it easy to exploit data-parallelism 
as required by many analysis applications. Several recent publications applied 
HPC techniques to advance traditional Big Data applications and Big Data frameworks
~\cite{fox-2017}.

Task-parallel applications involve partitioning a workload into a set of self-contained 
units of work. Based on the application, these tasks can be independent, have no 
inter-task communication, or coupled with varying degrees of data dependencies. 
Big Data applications exploit task parallelism for data-parallel parts (e.\,g., 
\texttt{map} operations), but also require coupling, for computing aggregates 
(the \texttt{reduce} operation). The MapReduce~\cite{mapreduce} abstraction 
popularized this execution pattern. Typically, a reduce operation includes shuffling 
intermediate data from a set of nodes to node(s) where the reduce executes. There 
is a recognized need to optimize communication intensive parts of Big Data frameworks 
using established HPC techniques for interprocess, e.\,g. shuffle operations~\cite{rdma-spark} 
and other forms of communication~\cite{hpc-abds,twisterNet}.

Spark~\cite{Zaharia_2010} and Dask~\cite{matthew_rocklin-proc-scipy-2015} are two 
Big Data frameworks. Both provide MapReduce abstractions and are optimized for 
parallel processing of large data volumes, interactive analytics 
and machine learning. Their runtime engines can automatically partition data, 
generate parallel tasks, and execute them on a cluster. In addition, Spark offers 
in-memory capabilities allowing caching data that are read multiple times, making 
it suited for interactive analytics and iterative machine learning algorithms. 
Dask also provides a MapReduce API (Dask Bags). Furthermore, Dask's API is more 
versatile, allowing custom workflows and parallel vector/matrix computations.

In this paper, we investigate the data analysis requirements of Molecular Dynamics 
(MD) applications. MD simulations are significant consumers of computing cycles, 
producing immense amounts of data. A typical $\mu sec$ MD simulation of physical 
system of $O(100k)$ atoms can produce from $O(10)$ to $O(1000)$ GBs of data~
\cite{Cheatham:2015}. In addition to being the prototypical HPC application, there 
is increasingly a need for the analysis to be integrated with simulations and 
drive the next stages of execution~\cite{extasy}. The analysis phase must be 
performed quickly and efficiently in order to steer the simulations.

We investigate three task-parallel frameworks and their suitability for implementing 
MD trajectory analysis algorithms. In addition to Spark and Dask, we investigate 
\rp~\cite{rp-jsspp18}, a Pilot-Job~\cite{pstar12} framework designed for implementing 
task-parallel applications on HPC. We utilize MPI4py~\cite{mpi4py_paper} to provide 
MPI equivalent implementations of the algorithms. The task-parallel implementations 
performance and scalability compared to MPI is the basis of our analysis. 
MD trajectories are time series of atoms/particles positions and velocities, which 
are analyzed using different statistical methods to infer certain properties, 
e.\,g. the relationship between distinct trajectories, snapshots of a trajectory 
etc. As a result, they can be considered as a representative set of scientific 
datasets that are organized as time series and their analysis algorithms. 

The paper makes the following contributions: 
\begin{inparaenum}[i)]
    \item it characterizes and explains the behavior of different MDAnalysis 
    algorithms on these frameworks, and
    \item provides a conceptual basis for comparing the abstraction, capabilities 
    and performance of these frameworks.
\end{inparaenum}

The paper is organized as follows: Section~\ref{use_cases} discusses the MD analysis 
algorithms investigated, and provides a brief characterization based on 
the Big Data Ogres classification ontology~\cite{bigdata-ogres}. Section~\ref{frameworks}, 
describes the different frameworks that were used for evaluation. 
Section~\ref{impl_exp} provides a description of the implementation of the MD 
algorithms on top of \rp, Spark and  Dask, as well as a performance evaluation 
and a discussion of findings. Section~\ref{related_work} reviews different MD 
analysis frameworks with respect to their ability to support scalable 
analytics of large volume MD trajectories. The paper concludes with a summary and 
discussion of future work in section~\ref{concl}.\up
\section{Molecular Dynamics Analysis Applications}
\label{use_cases}
Some commonly used algorithms for analyzing MD trajectories are Root Mean Square 
Deviation (RMSD), Pairwise Distances (PD), and Sub-setting~\cite{Mura:2014kx}. 
Two more advanced algorithms are Path Similarity Analysis (PSA)~\cite{10.1371/journal.pcbi.1004568} 
and Leaflet Identification~\cite{mdanalysis}. RMSD is used to identify the deviation 
of atom positions between frames. PD and PSA methods calculate distances based 
on different metrics either between atoms or trajectories. Sub-setting methods 
are used to isolate parts of interest of MD simulation. Leaflet Identification 
provides information about groups of lipids by identifying the lipid leaflets in 
a lipid bilayer. All these methods, in some way, read and process the whole physical 
system generated via simulations. The analysis reduces the data to either a 
number or a matrix.

We discuss two of these methods, a Path Similarity Analysis (PSA) algorithm using 
the Hausdorff distance and a Leaflet Identification method, and their implementations 
in MDAnalysis~\cite{mdanalysis,oliver_beckstein-proc-scipy-2016}. In addition, we 
implemented the PSA algorithm using CPPTraj~\cite{cpptraj-2013}. Furthermore, we 
explore the applications' Ogres Facets and Views~\cite{bigdata-ogres}, which provide 
a more systematic characterization.

Big Data Ogres~\cite{bigdata-ogres} are organized into four classes, called \emph{views}. 
The possible features of a view are called \emph{facets}. A combination of facets 
from all views defines an Ogre. The Views are: 
\begin{inparaenum}[1)]
    \item execution - describes aspects, such as I/O, memory, compute ratios, 
    whether computations are iterative, and the 5 V's of Big Data (Volume, 
    Velocity, Value, Variety and Veracity),
    \item data source \& style - discusses input data collection, storage and access,
    \item processing - describes algorithms and kernels used for computation, and
    \item problem architecture - describes the application architecture.
\end{inparaenum}

\up
\subsection{MDAnalysis}
\label{sec:mda}
MDAnalysis is a Python library~\cite{mdanalysis,oliver_beckstein-proc-scipy-2016} 
that provides a comprehensive environment for filtering, transforming and analyzing 
MD trajectories in all commonly used file formats. It provides a common object-oriented 
API to trajectory data and leverages existing libraries in the scientific Python 
software stack, such as NumPy~\cite{numpy} and Scipy~\cite{scipy}.

\upp
\subsubsection{Path Similarity Analysis (PSA): Hausdorff Distance}Path Similarity 
Analysis (PSA)~\cite{10.1371/journal.pcbi.1004568} is used to quantify the similarity 
between trajectories considering their full atomic detail. The basic idea is 
to compute pair-wise distances (for instance, using the Hausdorff metric~\cite{Huttenlocher:1993zr}) 
between members of an ensemble of trajectories and cluster the trajectories based 
on their distance matrix.

Each trajectory is represented as a two dimensional array. The first dimension 
corresponds to time frames of the trajectory, the second to the $N$ atom positions, 
in 3-dimensional space.\up

\begin{algorithm}[ht]
    \scriptsize
    \caption{Path Similarity Algorithm: Hausdorff Distance}
    \label{alg:hausdorff}
    \begin{algorithmic}[1]
        \Procedure{HausdorffDistance}{$T_1$,$T_2$}\Comment{$T_1$ and $T_2$ are a set of 
            3D points}
        \State \texttt{List $D_1$,$D_2$}
        \For{$\forall frame_1$ in $T_1$}
        \For{$\forall frame_2$ in $T_2$}
        \State \texttt{Append in $D_1$ $d_{RMS}$($frame_1$, $frame_2$)}
        \EndFor
        \State \texttt{$D_{t_1}$ append $min(D_1)$}
        \EndFor
        \For{$\forall frame_2$ in $T_2$}
        \For{$\forall frame_1$ in $T_1$}
        \State \texttt{Append in $D_2$ $d_{RMS}$($frame_2$, $frame_1$)}
        \EndFor
        \State\texttt{$D_{t_2}$ append $min(D_2)$}
        \EndFor
        \State \textbf{return} $max\Big(max(D_{t_1}),max(D_{t_2})\Big)$
        \EndProcedure
        \\        
        \Procedure{PSA}{$Traj$}\Comment{$Traj$ is a set of trajectories}
        \For{$\forall T_1$ in $Traj$}
        \For{$\forall T_2$ in $Traj$}
        \State \texttt{ $D_{( T_1,T_2 )}$=HausdorffDistance$\Big( T_1,T_2 \Big)$} 
        \EndFor
        \EndFor
        \State \Return $D$
        \EndProcedure
    \end{algorithmic}
\end{algorithm}

\up Algorithm~\ref{alg:hausdorff} describes the PSA algorithm with the Hausdorff metric 
over multiple trajectories. We apply a 2-dimensional data partitioning over the 
output matrix to parallelize, shown in algorithm~\ref{alg:partition}. Our Hausdorff 
metric calculation is based on a naive algorithm. Recently, an algorithm was 
introduced that uses early break to speedup execution~\cite{fasthausdorff}, although 
we are not aware of a parallel implementation of this algorithm.

\begin{algorithm}[ht]
    \scriptsize
    \caption{Two Dimensional Partitioning}
    \label{alg:partition}
    \begin{algorithmic}[1]
        \State Initially, there are $N^2$ distances, where $N$ is the number of trajectories. 
        Each distance defines a computation task.
        \State Map the initial set to a smaller set with $k=N/n_1$ elements, where $n_1$ is a 
        divisor of $N$, by grouping $n_1$ by $n_1$ elements together.
        \State Execute over the new set with $k^2$ tasks. Each task is 
        the comparisons between $n_1$ and $n_1$  elements of the initial set. They are executed 
        serially.
    \end{algorithmic}
\end{algorithm}

The algorithm is embarrassingly parallel and uses linear algebra kernels for 
calculations. It has complexity $O(n^2)$ (problem architecture \& processing views). 
Input data volume is medium to large while the output is small. Specific execution 
environments, such as HPC nodes, and Python arithmetic libraries, e.g., NumPy, are 
used (execution view). Input data are produced by HPC simulations, and are stored 
on HPC storage systems, such as parallel filesystem like Lustre (data source \& 
style view).\up

\subsubsection{Leaflet Finder}Algorithm~\ref{alg:leafletfinder} describes the 
Leaflet Finder (LF) algorithm as presented in Ref.~\cite{mdanalysis}. LF assigns 
particles to one of two curved but locally approximately parallel sheets, provided 
that the inter-particle distance is smaller than the distance between sheets. 
In biomolecular simulations of lipid membranes, consisting of a double layer of 
lipid molecules, LF is used to identify the lipids in the outer and inner leaflets 
(sheets). The algorithm consists of two stages:
\begin{inparaenum}[a)]
    \item construction of a graph connecting particles based on threshold distance 
    (cutoff), and
    \item computing the connected components of the graph, determining the lipids 
    located on the outer and inner leaflets.
\end{inparaenum}\up

\begin{algorithm}[ht]
    \scriptsize
    \caption{Leaflet Finder Algorithm}
    \label{alg:leafletfinder}
    \begin{algorithmic}[1]
        \Procedure{LeafletFinder}{$Atoms,Cutoff$}
        \Comment{$Atoms$ is a set of 3D points that represent the position of atoms in 
            space. $Cutoff$ is an Integer Number}
        \State \texttt{Graph G =$(V=Atoms,E=\emptyset)$}
        \For{$\forall atom$ in $Atoms$}
        \State \texttt{$N = [a\in V: d(a,atom)\le Cutoff]$}
        \State \texttt{Add edges $[(atoms,a): a \in N]$ in G}
        \EndFor
        \State \texttt{C = ConnectedComponents(G)}
        \State \Return C
        \EndProcedure
    \end{algorithmic}
\end{algorithm}
\up
The application stages have different complexities. The first stage identifies 
neighboring atoms. There are different implementation alternatives: 
\begin{inparaenum}[i)]
    \item computing the distance between all atoms ($O(n^2)$);
    \item utilizing a tree-based nearest neighbor (Construction: $O(n\log n)$, 
    Query: $O(\log n)$).
\end{inparaenum}
In both alternatives the input data volume is medium size and the output is smaller 
than the input. The complexity of connected components is: $O(|V|+|E|)$ ($V$: 
Vertices, $E$: Edges), i.\,e. it greatly depends on the characteristics of the 
graph.

The application typically uses HPC nodes as the execution environment, and NumPy 
arrays (execution view). It uses matrices to represent the physical system and 
the distance matrix. The output data representation is a graph. Leaflet Finder 
can be efficiently implemented using the MapReduce abstraction (problem architecture 
view). It uses graph algorithms and linear algebra kernels (processing view facets). 
The data source \& style view facets are the same as the PSA algorithm.\upp

\subsection{CPPTraj}
CPPTraj~\cite{cpptraj-2013,cpptraj_2018} is a C++ MD trajectory analysis tool. 
It is parallelized via MPI and OpenMP. CPPtraj reads in parallel frames from a 
single trajectory file or ensemble members of the same trajectory from different 
files. The frames are equally distributed to the MPI processes. Computational 
intensive algorithms are further parallelized using OpenMP. It requires at least 
one MPI process per ensemble member, where each member is a single trajectory file.
\section{Background of Evaluated Frameworks}
\label{frameworks}
The landscape of frameworks for data-intensive applications is manifold~\cite{JhaQLMF14,fox-2017} 
and has been extensively studied in the context of scientific~\cite{3dpas_intro} 
applications. In this section, we investigate the suitability of frameworks such 
as Spark~\cite{Zaharia_2010}, Dask~\cite{matthew_rocklin-proc-scipy-2015} and \rp
~\cite{rp-jsspp18}, for MD data analytics.

MapReduce~\cite{mapreduce} and its open source Hadoop implementation combined a 
simple functional API with a powerful distributed computing engine that exploits 
data locality to allow optimized I/O performance. However, MapReduce is inefficient 
for interactive workloads and iterative machine learning algorithms~\cite{Zaharia_2010,Ekanayake2010}. 
Spark and Dask provide richer APIs, caching and other capabilities critical for 
analytics applications. Spark is considered the standard solution for iterative 
data-parallel applications. Dask is quickly gaining support in the scientific 
community, since it offers a fully python environment. \rp is a Pilot-Job framework
~\cite{pstar12} that supports task-level parallelism on HPC resources. It successfully 
adds a parallelization level on top of HPC MPI-based applications.

\begin{figure}[ht]
    \centering
    \includegraphics[width=.48\textwidth]{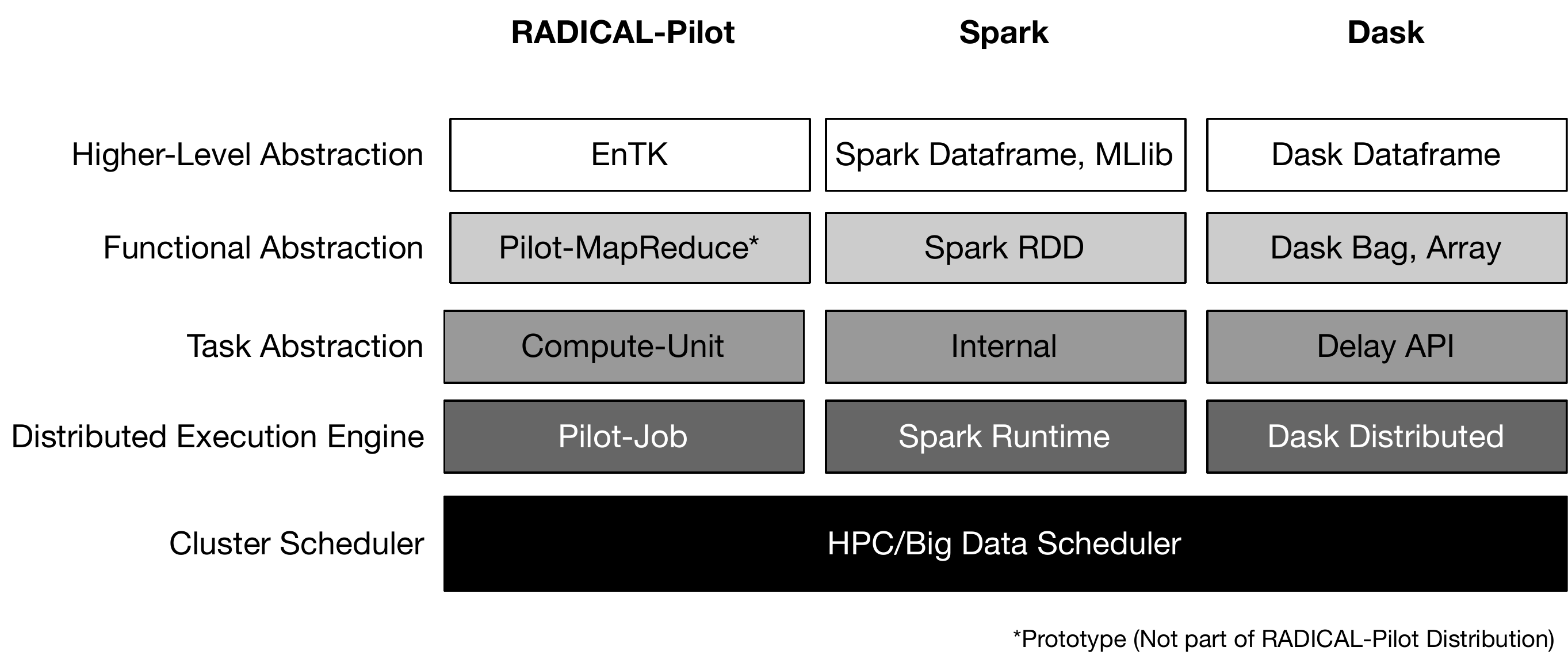}
    \caption{\textbf{Architecture of \rp, Spark and Dask:} The frameworks share 
        common architectural components for managing cluster resource, and tasks. 
        Spark, Dask offer several high-level abstractions inspired by MapReduce.\up}
    \label{fig:figures_bigdata_framework_stack}
\end{figure}

As described in~\cite{JhaQLMF14}, these frameworks typically comprise of distinct 
layers, e.\,g.,cluster scheduler access, framework-level scheduling, and 
higher-level abstractions. On top of low-level resource management capabilities 
various higher-level abstractions can be provided, e.\,g., MapReduce-inspired APIs. 
These then provide the foundation for analytics abstractions, such as Dataframes, 
Datasets and Arrays. Figure~\ref{fig:figures_bigdata_framework_stack} 
visualizes the different components of \rp, Spark and Dask. The following 
describes each framework in detail.\up

\up
\subsection{Spark}
Spark~\cite{Zaharia_2010} extends MapReduce~\cite{mapreduce} providing a rich set 
of operations on top of the Resilient Distributed Dataset (RDD) abstraction~\cite{Zaharia_RDD2012}. 
RDDs are cached in-memory making Spark well suitable for iterative applications 
that need to cache a set of data across multiple stages. PySpark provides a Python 
API to Spark.

A Spark job is compiled of multiple stages; a stage is a set of parallel tasks 
executed without communicating (e.\,g., \texttt{map}) and an action (e.\,g., 
\texttt{reduce}). Each action defines new stage. The \texttt{DAGScheduler} is 
responsible for translating the workflow specified via RDD transformations and 
actions to an execution plan. Spark's distributed execution engine handles the 
low-level details of task execution. The execution of a Spark job is triggered 
by actions. Spark can read data from different sources, such as HDFS, blob storage, 
parallel and local filesystems. While Spark caches loaded data in memory, it 
offloads to disk when an executor does not have enough free memory to hold all 
the data of its partition. Persisted RDDs remain in memory, unless specified to 
use the disk either complementary or as a single target. In addition, Spark writes 
to disk data that are used in a shuffle. As a result, it allows quick access to 
those data when transmitted to an executor. Finally, Spark provides a set of 
actions that write text files, Hadoop sequence files or object files to 
local filesystems, HDFS or any filesystem that supports Hadoop. In addition, 
Spark supports higher-level data abstractions for processing structured data, 
such as dataframes, Spark-SQL, datasets, and data streams.\up

\subsection{Dask}\upp
Dask~\cite{matthew_rocklin-proc-scipy-2015} provides a Python-based parallel computing 
library, which is designed to parallelize native Python code 
written for NumPy and Pandas. In contrast to Spark, Dask also provides a lower-level 
task API (\texttt{delayed} API) that allows users to construct arbitrary 
workflow graphs. Being written in Python, it does not require to translate data 
types from one language to another like PySpark, which moves 
data between Python's interpreter and Java/Scala. 

In addition to the low-level task API, Dask offers three higher-level abstractions: 
Bags, Arrays and Dataframes. Dask Arrays are a collection of NumPy arrays 
organized as a grid. Dask Bags are similar to Spark RDDs and are used to analyze 
semi-structured data, like JSON files. Dask Dataframe is a distributed collection 
of Pandas dataframes that can be analyzed in parallel.

Furthermore, Dask offers three schedulers: multithreading, multiprocessing and 
distributed. The multithreaded and multiprocessing schedulers can be used only on 
a single node and the parallel execution is done via threads and processes 
respectively. The distributed scheduler creates a cluster with a scheduling process 
and multiple worker processes. A client process creates and communicates a DAG to 
the scheduler. The scheduler assigns tasks to workers.

Dask's learning curve cannot be considered steep. Its API is well defined and 
documented. In addition, familiarity with Spark or MapReduce helps to minimize 
the learning curve even further. As a result, implementing MD analysis algorithms 
on Dask did not require significant engineering time. In addition, setting up a 
Dask cluster on a set of resources was relatively straightforward, since it provides 
all the binaries, e.g. \texttt(dask-ssh).\up

\subsection{RADICAL-Pilot}
\upp\rp~\cite{rp-jsspp18} is a Pilot system that implements the pilot paradigm 
as outlined in Ref.~\cite{turilli2017comprehensive}. \rp (RP) is implemented in Python 
and provides a well defined API and usage modes. Although RP is vehicle for research 
in scalable computing, it also supports production grade science. Currently, it 
is being used by applications drawn from diverse domains, ranging from earth and 
biomolecular sciences to high-energy physics. RP can be used as a 
runtime system by workflow or workload management systems~\cite{bb_2016,repex2016,power-of-many17,dakka2017htbac,turilli2016analysis}. 
In 2017, RP was used to support more than 100M core-hours on US DOE, NSF resources 
(BlueWaters and XSEDE), and European supercomputers (Archer and SuperMUC).

\rp allows concurrent task execution on HPC resources. The user defines a set of 
\computeunits (\cu)~- the abstraction that defines a task along with its dependencies 
- which are submitted to \rp. \rp schedules these \cus to be executed under the 
acquired resources. It uses the existing environment of the resource to execute 
tasks. Any data communication between tasks is done via an underlying 
shared filesystem, e.g., Lustre. Task execution coordination and communication 
is done through a database (MongoDB).

\rp's learning curve can be quite steep at the beginning, at least until the user 
becomes familiar with the concept and usability of Pilots and \cus. Once the user 
is comfortable with \rp's API, she can easily develop new algorithms.\up

\subsection{Comparison of Frameworks}
\begin{table}[t]
    \scriptsize
    \begin{tabular}{|p{1.4cm}|p{1.8cm}|p{1.8cm}|p{1.8cm}|}
        \hline
        &\textbf{RADICAL-Pilot} &\textbf{Spark} &\textbf{Dask} \\ \hline
        Languages       &Python      &Java, Scala, Python, R       &Python\\\hline
        Task &Task&Map-Task&Delayed\\
        Abstraction         &      &       & \\\hline
        Functional Abstraction  &-         &RDD API       &Bag\\\hline        
        Higher-Level Abstractions &EnTK~\cite{power-of-many17} &Dataframe, ML 
        Pipeline, MLlib~\cite{JMLR:v17:15-237} &Dataframe, Arrays for block computations\\\hline
        Resource Management &Pilot-Job &Spark Execution Engines &Dask 
        Distributed Scheduler\\\hline
        Scheduler    &Individual Tasks &Stage-oriented DAG &DAG\\\hline        
        Shuffle      &-       &hash/sort-based shuffle &hash/sort-based shuffle
        \\ \hline
        Limitations &no shuffle, filesystem-based communication  &high overheads for Python tasks (serialization)   & Dask Array can not deal with dynamic output shapes\\\hline
    \end{tabular}
    \caption{\textbf{Frameworks Comparison:} Dask and Spark are designed for data-related task, while \rp focuses on compute-intensive tasks.\label{tab:frameworks}}\up
\end{table}

Table~\ref{tab:frameworks} summarizes the properties of these frameworks with respect 
to abstractions and runtime provided to create and execute parallel 
data applications. 
\uph
\subsubsection*{API and Abstractions} 
\rp provides a low-level API for executing tasks onto resources. While this API 
can be used to implement high-level capabilities, e.\,g. MapReduce~\cite{pilot-mapreduce2012}, 
they are not provided out-of-the box. Both Spark and Dask provide such capabilities. 
Dask's API is generally lower level than Spark's , e.\,g., it allows specifying 
arbitrary task graphs. Although, data partition size is automatically decided, 
in many cases it is necessary to tune parallelism by specifying the number of 
partitions.

Another important aspect is the availability of high-level abstractions. High-level 
tools for \rp, such as Ensemble Toolkit~\cite{power-of-many17}, are designed 
for workflows involving compute-intensive tasks. Spark and Dask already offer a 
set of high-level data-oriented abstractions, such as Dataframes.
\uph
\subsubsection*{Scheduling}
Both Spark and Dask create a Direct Acyclic Graph (DAG) based on operations over 
data, which is then executed using their execution engine. Spark jobs are separated 
into stages. When a stage is completed, the scheduler executes the next stage.

Dask's DAGs are represented by a tree where each node is a task. Leaf tasks do 
not depend on other task for execution. Dask tasks are executed when their 
dependencies are satisfied, starting from leaf tasks. When a task is reached with 
unsatisfied dependencies, the scheduler executes the dependent task first. Dask's 
scheduler does not rely on synchronization points that Spark's stage-oriented 
scheduler introduces. \rp does provide a DAG and requires the execution order 
and synchronization to be specified by the user.
\uph
\subsubsection*{Suitability for MDAnalysis Algorithms}
Trajectory analysis methods are often embarrassingly parallel. So, they are ideally suited 
for task management and MapReduce APIs. PSA-like methods typically 
require a single pass over the data and return a set of values that correspond to 
a relationship between frames or trajectories. They can be expressed as a bag of 
tasks using a task management API or a map-only application in a MapReduce-style 
API. 

Leaflet Finder is more complex and requires two stages:
\begin{inparaenum}[a)]
    \item the edge discovery stage, and
    \item the connected components stage.
\end{inparaenum}
It is possible to implement Leaflet Finder with a simple task-management API, 
although the MapReduce programming model allows more efficient implementation 
with a \texttt{map} for computing and filtering distances and a \texttt{reduce} 
for finding the components. The shuffling required between map and reduce is medium 
as the number of edges is a fraction of the input data.\upp
\section{Experiments and Discussion}
\label{impl_exp}
In this section, we characterize the performance of \rp, Spark and Dask compared 
to MPI4py. In section~\ref{sec:framework_eval} we evaluate the task throughput 
using a synthetic workload. In sections~\ref{sec:psa} and~\ref{sec:leaflet} we 
evaluate the performance of two algorithms from MDAnalysis: PSA and Leaflet Finder 
using different real-world datasets. We investigate: 
\begin{inparaenum}[1)]
    \item which capabilities and abstractions of the frameworks are needed to 
    efficiently express these algorithms,
    \item what architectural approaches can be used to implement these algorithms 
    with these frameworks, and
    \item the performance trade-offs of these frameworks.
\end{inparaenum}

The experiments were executed on the XSEDE Supercomputers: Comet and Wrangler. 
SDSC Comet is a 2.7 PFlop/s cluster with 24\, Haswell cores/node and 128\,GB 
memory/node (6,400 nodes). TACC Wrangler has 24\, Haswell hyper-threading enabled 
cores/node and 128\,GB memory/node (120 nodes). Experiments were carried using \rp and Pilot-Spark~\cite{hadoop-on-hpc} 
extension, which allows to efficiently manage Spark on HPC resources through a 
common resource management API. We utilize a set of custom scripts to start the 
Dask cluster. We used \rp 0.46.3, Spark 2.2.0, Dask 0.14.1 and Distributed 1.16.3. 
The data presented are means over multiple runs; error bars represent the standard 
deviation of the sample. We employed up to $10$ nodes in Comet and Wrangler. 
\up
\subsection{Frameworks Evaluation}
\label{sec:framework_eval}

\begin{figure}[t]
    \centering
    \includegraphics[width=.48\textwidth]{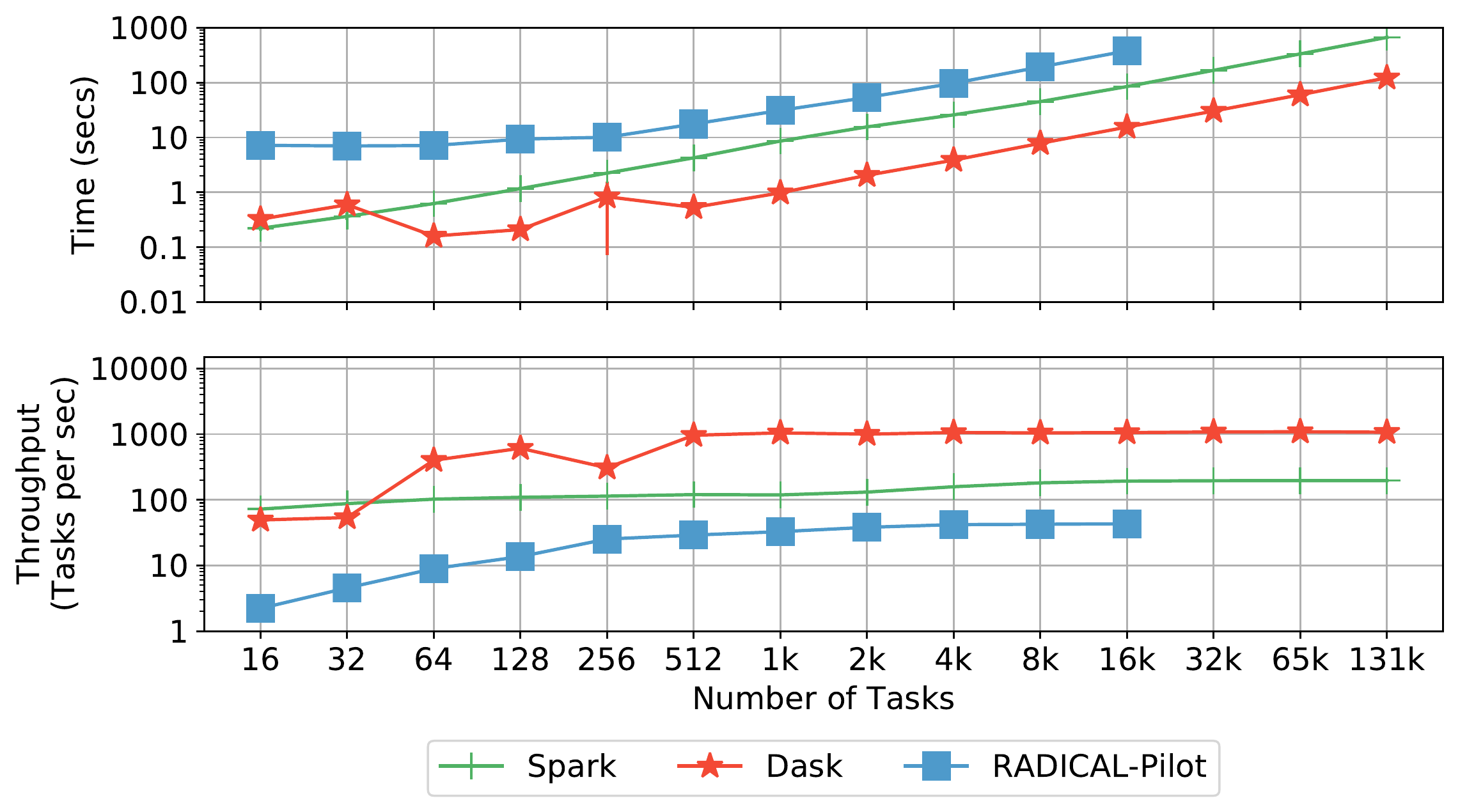}
    \caption{\textbf{Task Throughput by Framework (Single Node):} 
        Time/Throughput executing a given number of zero-workload tasks 
        on Wrangler. Dask performs best; Dask and Spark have very small delays for 
        few tasks. \rp offers the smallest throughput.\uph}
    \label{fig:dask_spark_rp_wrangler}
\end{figure}

As data-parallelism often involves a large number of short-running tasks, task 
throughput is a critical metric to assess the different frameworks. To evaluate 
the throughput we use zero workload tasks (\texttt{/bin/hostname}). We submit an 
increasing number of such tasks to \rp, Spark and Dask and measure the execution 
time.

For \rp, all tasks were submitted simultaneously. \rp's backend database was running 
on the same node to avoid large communication latencies. For Spark, we created 
an RDD with as many partitions as the number of tasks -- each partition is mapped 
to a task by Spark. For Dask, we created tasks using \texttt{delayed} functions 
that were executed by the Distributed scheduler. We used Wrangler and Comet for 
this experiment.

Figure~\ref{fig:dask_spark_rp_wrangler} shows the results. Dask needed the least 
time to schedule and execute the assigned tasks, followed by Spark and \rp. Dask 
and Spark quickly reach their maximum throughput, which is sustained as the 
number of tasks increased. \rp showed the worst throughput and scalability 
mainly due to some architectural limitations. It relies on a MongoDB to communicate 
between Client and Agent, as well as several components that allow \rp to move 
data and introduce delays in the execution of the tasks. Thus, we were not able to 
scale \rp to 32k or more tasks.

\begin{figure}[t]
    \centering
    \includegraphics[width=.48\textwidth]{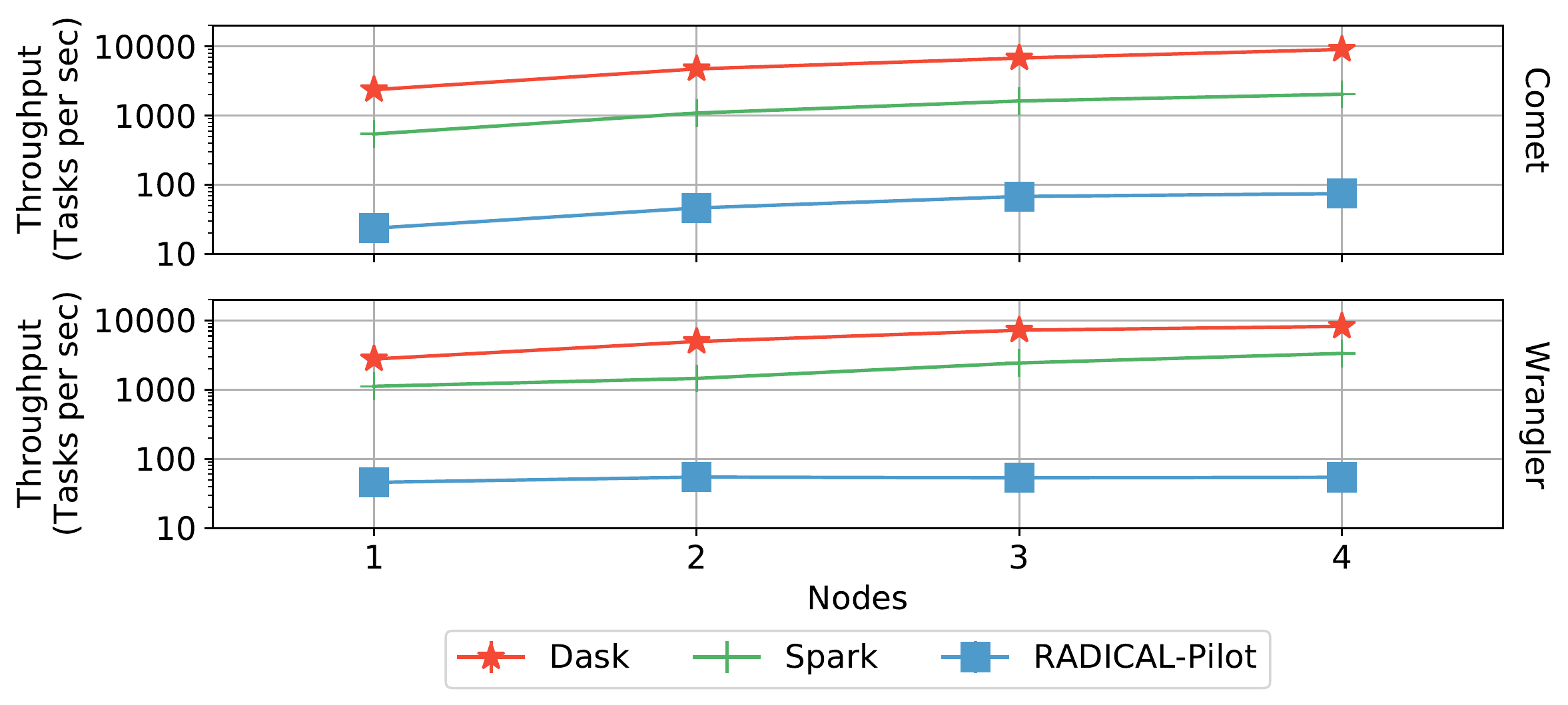}
    \caption{\textbf{Task Throughput by Framework (Multiple Nodes):} Task throughput for $100k$ 
        zero-workload tasks on different numbers of nodes for each framework. Dask has the largest 
        throughput, followed by Spark and \rp.\up}
    \label{fig:RP_Dask_Spark_throughput}
\end{figure}

Figure~\ref{fig:RP_Dask_Spark_throughput} illustrates the throughput when scaling 
to multiple nodes measured by submitting $100k$ tasks. Dask's throughput on all 
three resources increases almost linearly to the number of nodes. Spark's throughput 
is an order of magnitude lower than Dask's. \rp 's throughput plateaus at below 
$100 task/sec$. Wrangler and Comet show a comparable performance 
with Comet slightly outperforming Wrangler.\up

\subsection{Path Similarity Analysis: Hausdorff Distance}
\label{sec:psa}
The PSA algorithm is embarrassingly parallel and can be implemented using simple 
task-level parallelism or a map only MapReduce application. The input data, i.\,e. 
a set of trajectory files, is equally distributed over the cores, generating one 
task per core. Each task reads its respective input files in parallel, executes 
and writes the result to a file.

For \rp we define a \computeunit for each task and execute them using a Pilot-Job. 
For  Spark, we create an RDD with one partition per task. The tasks are executed 
in a \texttt{map} function. In Dask, the tasks are defined as \texttt{delayed} 
functions. In MPI, each task is executed by a process.

The experiments were executed on Comet and Wrangler. The dataset used consists 
of three different atom count trajectories: small ($3341$ atoms/frame), medium 
($6682$ atoms/frame) and large ($13364$ atoms/frame), and $102$ frames. We used 
$128$ and $256$ trajectories of each size. 

\begin{figure}[ht]
    \centering
    \includegraphics[width=0.48\textwidth]{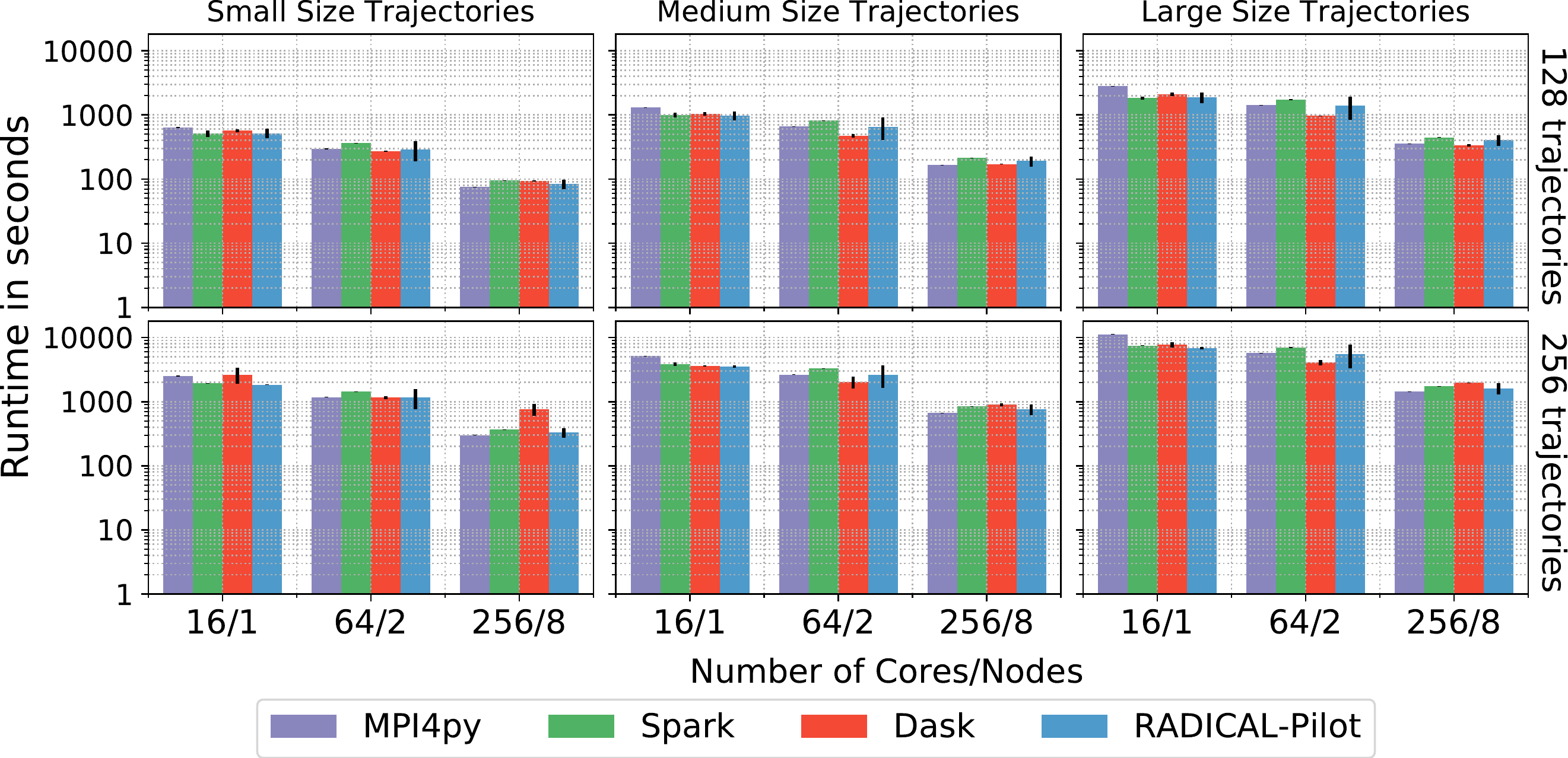}
    \caption{\label{fig:HausdorffWrangler}\textbf{Hausdorff Distance on Wrangler using 
            \rp, Spark and Dask:} Runtimes over different number of cores, 
        trajectory sizes, and number of trajectories. All frameworks scaled by a 
        factor of 6 from 16 to 256 cores.\up} 
\end{figure}

\begin{figure}[t]
    \centering
    \includegraphics[width=.48\textwidth]{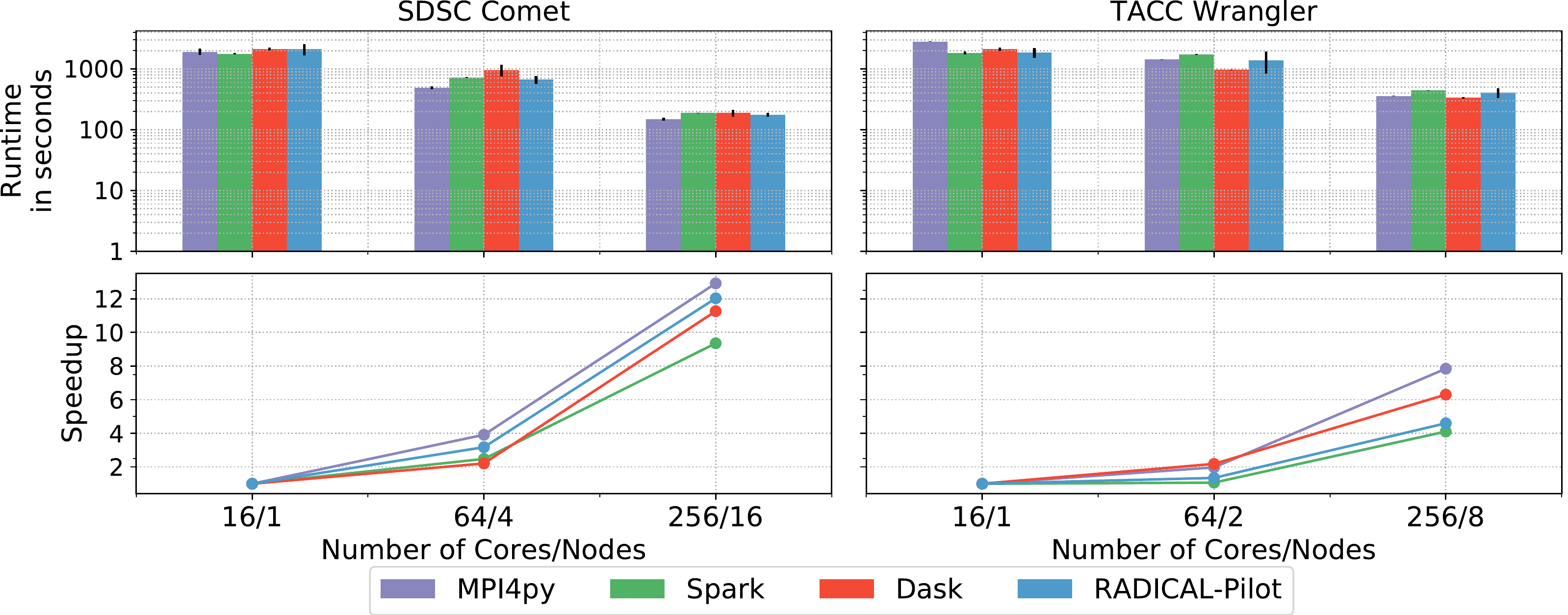}
    \caption{\textbf{Hausdorff Distance on Comet and Wrangler:}
        Runtime and Speedup for 128 large trajectories.\upp} 
\label{fig:comet_wrangler_haus}
\end{figure}

Figure~\ref{fig:HausdorffWrangler} shows the runtime for 128 and 256 trajectories 
on Wrangler. Figure~\ref{fig:comet_wrangler_haus} compares the execution times on 
Comet and Wrangler for $128$ large trajectories. We see that the frameworks have 
similar performance on both systems. Furthermore, we see that Wrangler gives 
smaller speedup than Comet. Although, we used the same number of cores, 
we see that utilizing half the nodes due to hyperthreading results to smaller 
speedup.

MPI4py, \rp, Spark and Dask have similar performance when used to execute embarrassingly 
parallel algorithms. All frameworks achieved similar speedups as the number of 
cores increased, which are lower than MPI4py. Although, the frameworks' overheads 
are comparably low in relation to the overall runtime, they were significant to 
impact their speedup. \rp's large deviation is due to sensitivity to communication 
delays with the database. In summary, all three frameworks provide appropriate 
abstractions and runtime performance, compared to MPI, for embarrassingly parallel 
algorithms. In this case aspects such as programmability and integrate-ability 
are more important considerations,e.\,g., both \rp and Dask are native Python 
frameworks making the integration with MDAnalysis easier and more efficient than 
with other frameworks, which are based on other languages.

\begin{figure}[ht]
    \centering
    \includegraphics[width=0.48\textwidth]{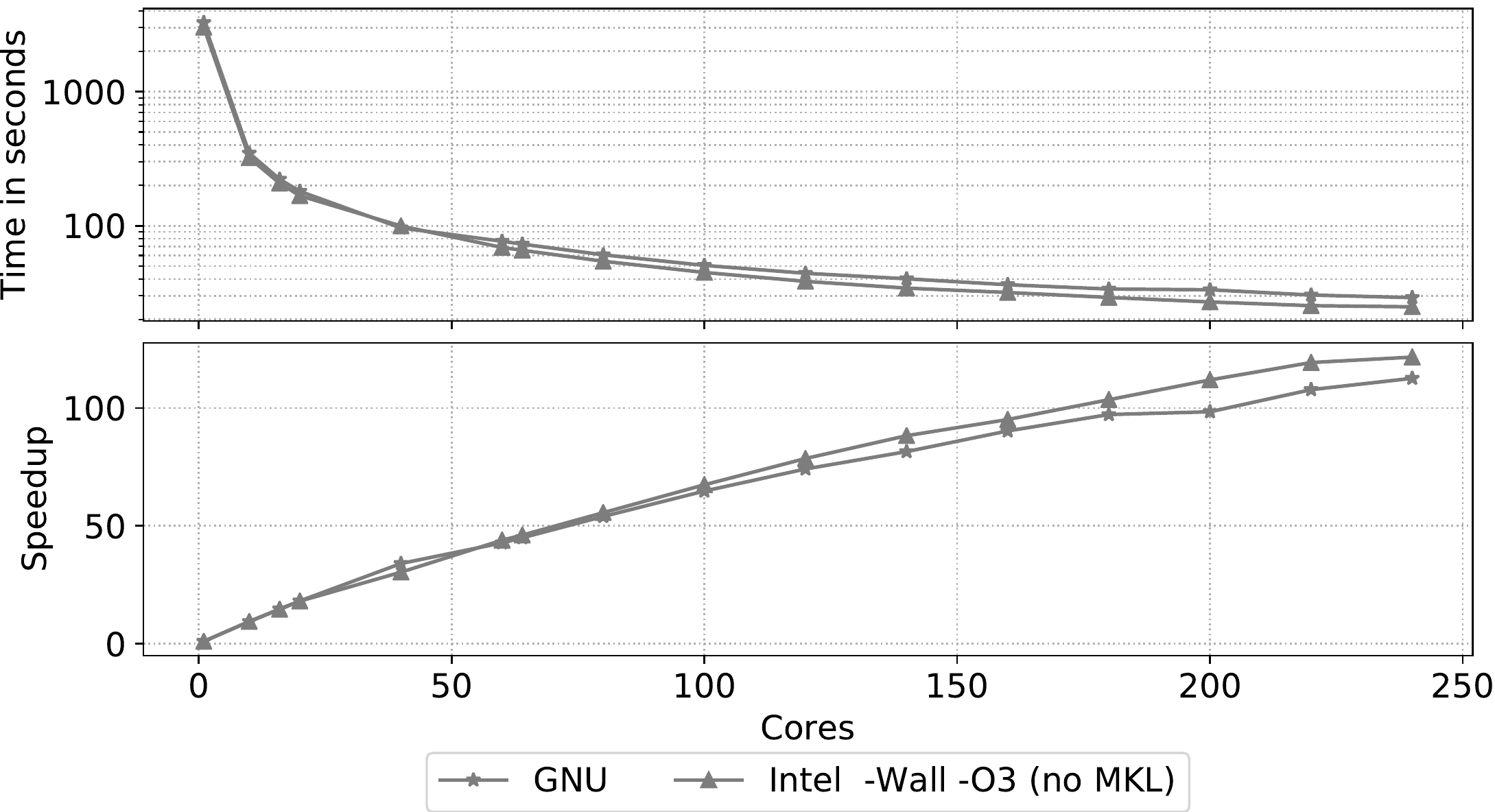}
    \caption{\label{fig:cpptraj_resutls}\textbf{Hausdorff Distance using 
            CPPTraj:} Runtimes and Speedup over different number of cores,\up} 
\end{figure}

CPPTraj~\cite{cpptraj_2018} provides an optimized C++ implementation 
of the 2D-RMSD, which is Algorithm~\ref{alg:hausdorff} with no $\min-\max$ operations. 
The 2D-RMSD between trajectories was executed in parallel. The results were
gathered and the Hausdorff distance was calculated. CPPTraj~\cite{cpptraj_2018} 
was compiled with GNU C++ compiler and no optimizations, and with Intel's compiler O3 
optimization enabled. An experiment was run with 20-core Haswell nodes and 128 small 
trajectories; number of cores ranging from 1 up to 240. Figure~\ref{fig:cpptraj_resutls} 
shows the runtimes and speedup. MPI C++ provides lower execution times. 
However, we are interested in scalable solutions, that may offer worse performance 
in absolute numbers, but allows easier integration, i.e., less lines of code, and/or 
less engineering time.\up

\subsection{Leaflet Finder}
\label{sec:leaflet}
\begin{table*}[ht]
	\centering
	\scriptsize
	\begin{tabular}{|p{1.65cm}|p{3cm}|p{3cm}|p{4.3cm}|p{3.7cm}|}
		\hline  
		&\textbf{Broadcast and 1-D} (Approach 1) &\textbf{Task API and 2-D} (Approach 2) &\textbf{Parallel Connected Components} (Approach 3) &\textbf{Tree-Search} (Approach 4)\\\hline
		Data Partitioning  & 1D  & 2D & 2D &2D\\ \hline
		
		Map &Edge Discovery via Pairwise Distance &Edge Discovery via Pairwise Distance &Edge Discovery via Pairwise Distance and Partial Connected Components 
		&Edge Discovery via Tree-based Algorithm and Partial Connected Components
		\\\hline
		
		Shuffle  &Edge List ($O(E)$) &Edge List ($O(E)$) &Partial Connected components ($O(n)$) &Partial Connected components ($O(n)$)\\\hline
		
		Reduce   &Connect Components  & Connected Components & Joined Connected 
		Components & Joined Connected Components\\\hline
	\end{tabular}
	\caption{MapReduce Operations used by Leaflet Finder\label{tab:app_operators}\up}
\end{table*}

In this section, we investigate four different approaches for implementing the 
Leaflet Finder algorithm using \rp, Spark, Dask, and MPI4py
(see Table~\ref{tab:app_operators}):
\begin{enumerate}[1)]
    \item \textbf{Broadcast and 1-D Partitioning:} The physical system is broadcast 
    and partitioned through a data abstraction. Use of RDD API (broadcast), Dask 
    Bag API (scatter), and MPI Bcast to distribute data to all nodes. A \texttt{map} 
    function calculates the edge list using \texttt{cdist} from SciPy~\cite{scipy} 
    -- realized as a loop for MPI. The list is collected to the master process 
    (gathered to rank 0) and the connected components are calculated.\label{en:1}
    \item \textbf{Task API and 2-D Partitioning:} Data management is done without 
    using the data-parallel API. The framework is used for task scheduling. Data 
    are pre-partitioned in 2-D partitions and passed to a \texttt{map} function 
    that calculates the edge list using \texttt{cdist}-- realized as a loop for 
    MPI. The list is collected (gathered to rank 0) and the connected components 
    are calculated.\label{en:2}
    \item \textbf{Parallel Connected Components:} Data are managed as in approach
    ~\ref{en:2}. Each \texttt{map} task performs edge list and connected components 
    computations. The reduce phase joins the calculated components into one, when 
    there is at least one common node.\label{en:3}
    \item \textbf{Tree-based Nearest Neighbor and Parallel-Connected Components 
        (Tree-Search):} This approach is different to approach~\ref{en:3} only on the 
    way edge discovery in the \texttt{map} phase is implemented. A tree containing 
    all atoms is created which is then used to query for adjacent atoms.\label{en:4}
\end{enumerate}

We use four physical systems with $131k$, $262k$, $524k$, and $4M$ atoms with 
$896k$, $1.75M$, $3.52M$, and $44.6M$ edges in their graphs. Experimentation was 
conducted on Wrangler where we utilized up to 256 cores. Data partitioning results 
into $1024$ partitions for each approach, thus $1024$ \texttt{map} tasks. Due to 
memory limitations from using \texttt{cdist} -- uses double precision floating 
point -- Approach \ref{en:3} data partitioning of the $4M$ atom dataset resulted 
to $42k$ tasks for both Spark and MPI4py.

\begin{figure*}[ht]
    \upp
    \begin{subfigure}{.48\textwidth}
        \centering
        \includegraphics[width=1\linewidth]{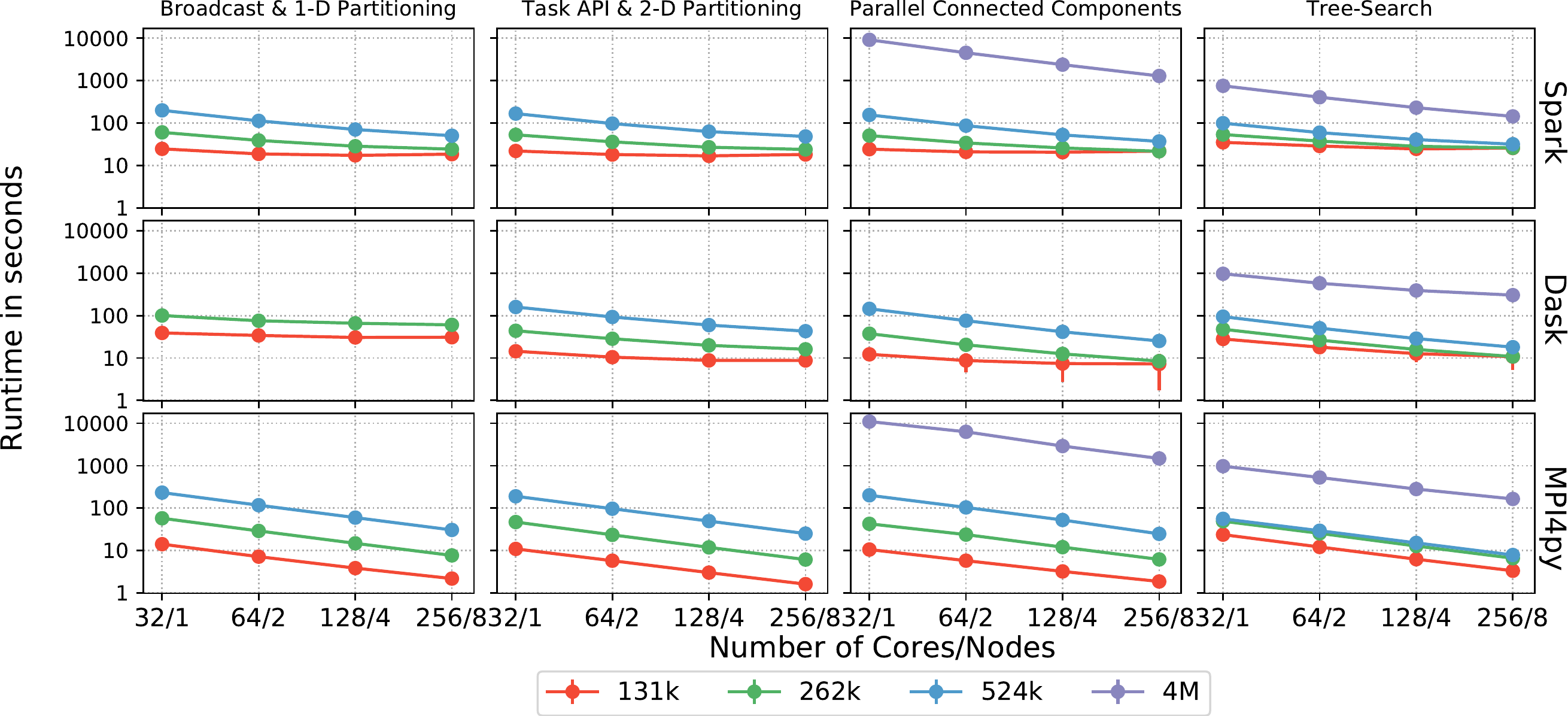}
    \end{subfigure}\upp%
    \begin{subfigure}{.48\textwidth}
        \centering
        \includegraphics[width=.95\linewidth]{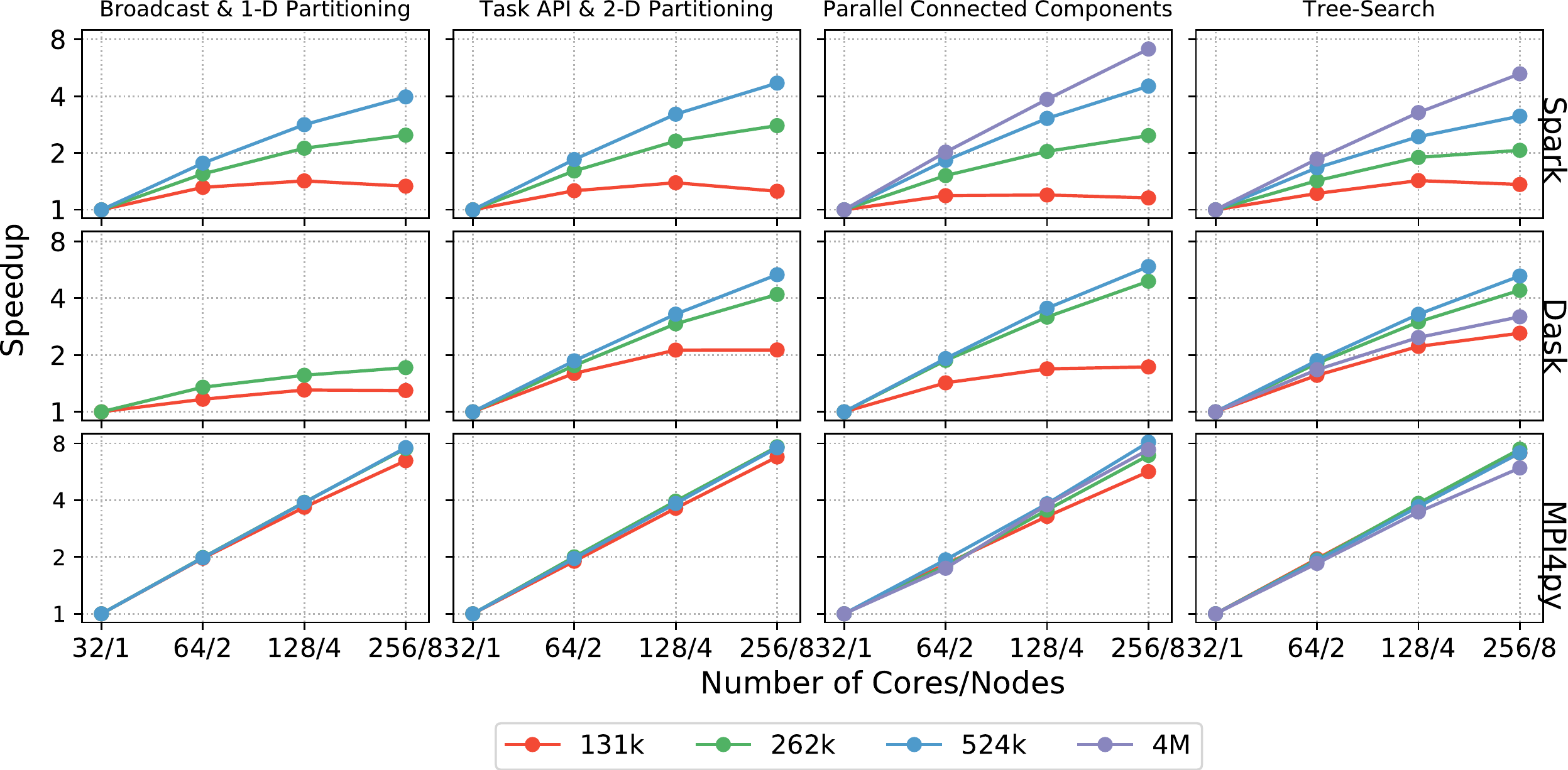}
    \end{subfigure}\upp
    \caption{\textbf{Leaflet Finder: Performance of Different
            Architectural Approaches for Spark \& Dask:} Runtimes and Speedups for 
        different system sizes over different number of cores for all approaches 
        and frameworks.\upp}
    \label{fig:All4approachesNoRp}
\end{figure*}

\begin{figure}[ht]
    \centering
    \includegraphics[width=.48\textwidth]{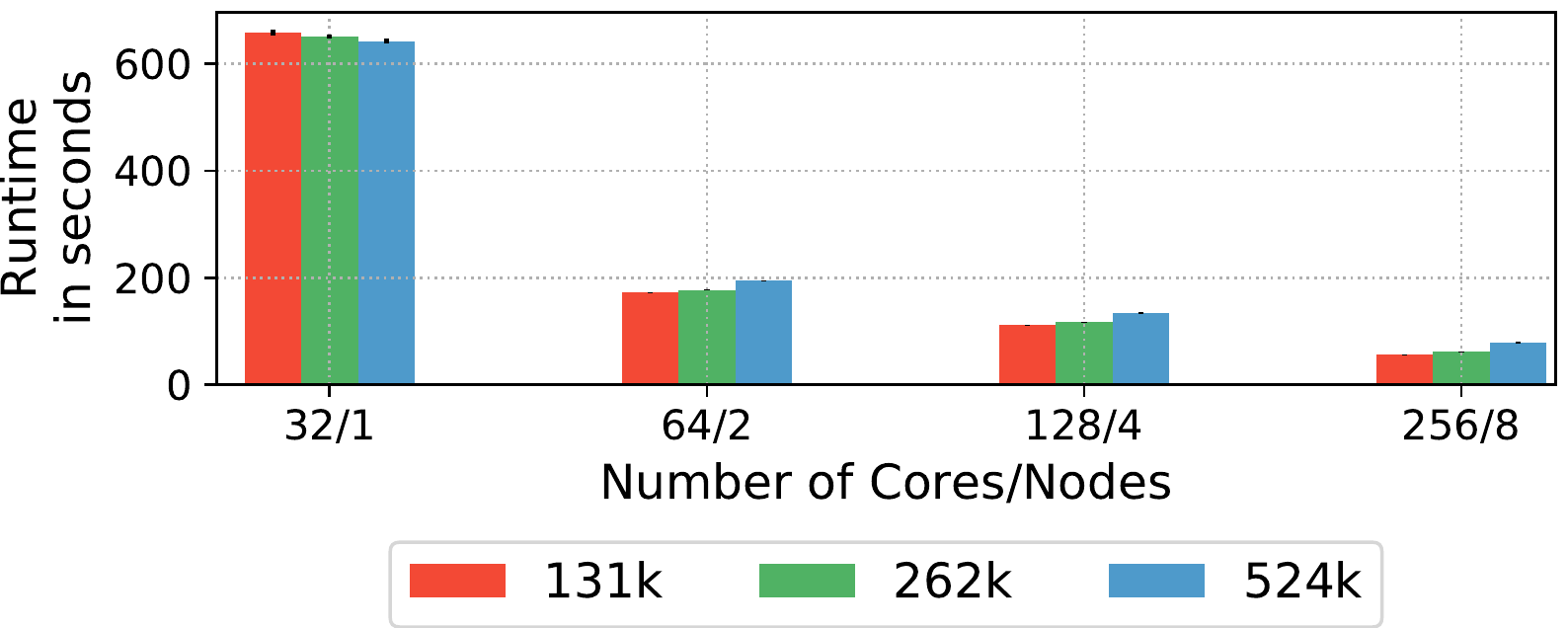}
    \caption{\textbf{\rp Task API and 2-D Partitioned Leaflet Finder
            (Approach 2):} Runtime for 
        multiple system sizes over different number of cores. Overheads dominate 
        since execution times are similar despite the system size.\uph}
    \label{fig:rpLF}
\end{figure}

Figure \ref{fig:All4approachesNoRp} shows the runtimes for all datasets for Spark, 
Dask and MPI4py. \rp's performance is illustrated in Figure~\ref{fig:rpLF}. We 
continue by analyzing the performance of each architectural approach and used 
framework in detail.

\begin{figure}[ht]
    \upp
    \centering
    \includegraphics[width=.48\textwidth]{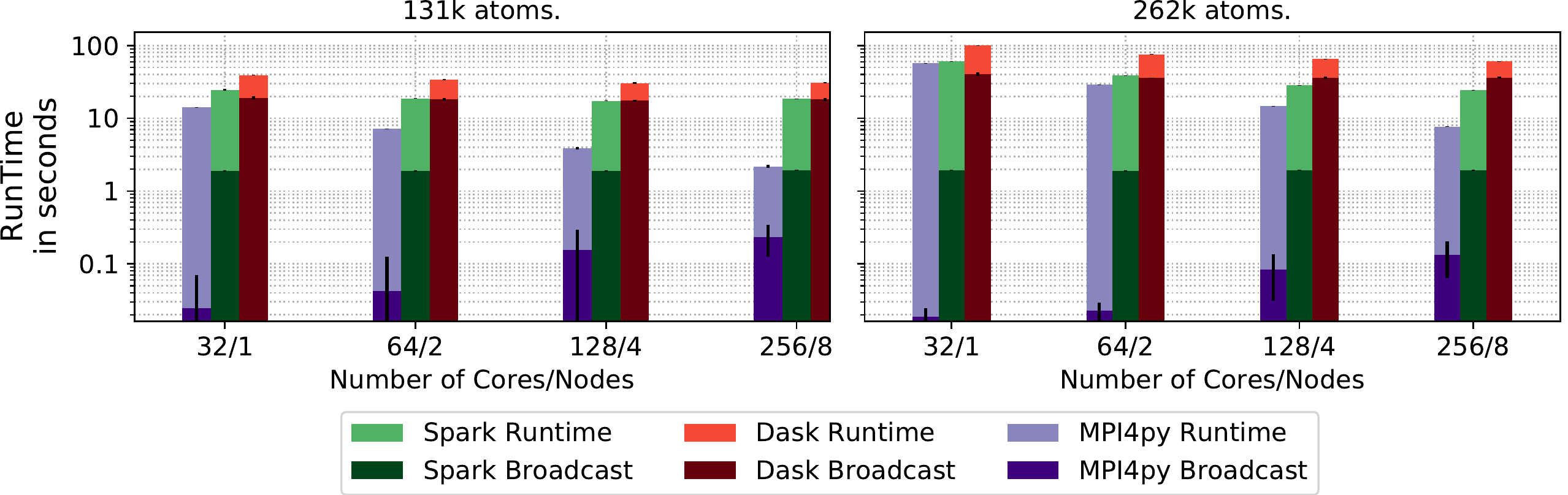}
    \caption{\textbf{Broadcast and 1-D Partitioned Leaflet Finder
            (Approach 1):} Runtime for multiple system sizes on different number of cores for 
        Spark, Dask and MPI4py.\upe}
    \label{fig:WranglerLeafLetFinderApp1}
\end{figure}
\upp
\subsubsection{Broadcast and 1-D Partitioning}
Approach 1 utilizes a broadcast to distribute the data to all nodes, which is 
supported by Spark, Dask and MPI. All nodes maintain a complete copy of the dataset. 
Each \texttt{map} task computes the pairwise distance on its partition. We use 
1-D partitioning. Figure~\ref{fig:WranglerLeafLetFinderApp1} shows the detailed 
results: as expected the usage of a broadcast has severe limitations for Spark and 
Dask. MPI broadcast is a fraction of the overall execution time and significantly 
smaller than Spark and Dask. MPI's broadcast times increase linearly as the number 
of processes increases, while Spark's and Dask's remain relatively constant for 
each dataset, due to more elaborate broadcast algorithms compared to MPI. Broadcast 
times are about $3\%$ -- $15\%$ of the edge discovery time for Spark, $40\%$ -- $65\%$ 
for Dask, and $<1\%$ -- $10\%$ for MPI4py. Spark offers a more efficient communication 
subsystem compared to Dask. In addition, Dask broadcast partitions the dataset 
to a list where each element represents a value from the initial dataset. This 
did not allow broadcasting the $524k$ atom dataset. Nevertheless, the 
limited scalability of this approach due to transmitting the entire dataset 
renders it only usable for small datasets. It shows the worst performance and scaling 
of all approaches for Spark, Dask and MPI4py.

Furthermore, this approach only scales up to $262k$ atoms for Dask, and $524k$ atoms 
for Spark and MPI4py on Wrangler. Spark's performance is comparable to MPI4py for 
the $262k$, and $524k$ datasets. It also shows better performance for the smallest 
core count in the $524k$ case. Dask is at least two times slower than our MPI 
implementation.


\upp
\uph
\subsubsection{Task-API and 2-D Partitioning}
Approach~\ref{en:2} tries to overcome the limitations of approach 1, especially 
broadcasting and 1-D partitioning. A 2-D block partitioning is essential, as it 
evenly distributes the compute and more efficiently utilizes the available memory. 
2-D partitioning is not well supported by Spark and Dask. Spark's RDDs are optimized 
for data-parallel applications with 1-D partitioning. While Dask's array supports 
2-D block partitioning, it was not used for this implementation. We return the 
adjacency list of the graph instead of an array to fully use the capabilities of 
the abstraction. Thus, each task works on a 2-D pre-partitioned part of the input 
data.

Figure~\ref{fig:All4approachesNoRp} shows the runtimes of approach~\ref{en:2} for 
Spark, Dask, MPI4py and Figure~\ref{fig:rpLF} for \rp. As expected this approach 
overcomes the limitations of approach 1 and can easily scale to larger datasets 
(e.\,g., $524k$ atoms) while improving the overall runtime. Dask's execution time 
was smaller by at least a factor of two. However, we were not able to scale this 
implementation to the 4M dataset, due to memory requirements of \texttt{cdist}. 
For \rp we observed significant task management overheads (see also section
~\ref{sec:framework_eval}). This is a limitation of \rp with respect to managing 
large numbers of tasks. This is particularly visible when the scenario was run on 
a single node with 32 cores. As more resources become available, i.e. more than 
64 cores, the performance improves dramatically.

Furthermore, Spark and Dask did not scale as well as MPI, which achieved linear 
speedups of $\sim8$ when using $256$ cores. Spark and Dask achieved maximum speedups 
of $4.5$ and $\sim5$ respectively. Despite this fact, both frameworks had similar 
performance on $32$ cores for the $262k$ and $524k$ datasets.\up


\subsubsection{Parallel Connected Components}
Communication between the edge discovery and connected components stages is another 
important aspect. The edge discovery phase output for the $524k$ atoms dataset is 
$\approx$$100\,\textup{MB}$. To reduce the amount of data that need to be shuffled, 
we refined the algorithm to compute the graph components on the partial dataset 
in the \texttt{map} phase. The partial components are then merged in a \texttt{reduce} 
phase. This reduces the amount of shuffle data by more than $50\%$ (e.\,g., to 
$12\textup{MB}$ for Spark and $48\textup{MB}$ for Dask). Figure~\ref{fig:All4approachesNoRp} 
shows the improvements in runtime, by $\sim20\%$ for Spark and Dask, but not 
MPI4py. Further, we were able to run very large datasets, such as the 4M dataset, 
using this architectural approach using Spark and MPI4py. Dask was restarting its 
worker processes because their memory utilization was reaching $95\%$.

Spark, and Dask have comparable performance with MPI on 32 cores, which utilizes 
a single node on Wrangler. However, the MPI4py implementation scales almost linearly 
for all datasets, Spark and Dask cannot, reaching a maximum of $\sim5$ for the 
three smaller datasets. In addition, Spark is able to scale almost linearly for 
the $4M$ atoms dataset providing comparable performance to MPI4py.

\up
\subsubsection{Tree-Search}
A bottleneck of approaches~\ref{en:1},~\ref{en:2} and~\ref{en:3} is the edge 
discovery via the naive calculation of the distances between all pairs of atoms. 
In approach~\ref{en:4}, we replace the pairwise distance function with a tree-based, 
nearest neighbor search algorithm, in particular BallTree~\cite{Omohundro89fiveballtree}. 
The algorithm: 
\begin{inparaenum}
    \item constructs of a tree, and
    \item queries for neighboring atoms.
\end{inparaenum}
Using tree-search, the computational complexity can be reduced from $n^2$ to $log$. 
We use a BallTree as offered by Scikit-Learn~\cite{scikit-nearest} for our 
implementation.

Figure \ref{fig:All4approachesNoRp} illustrates the performance of the implementation. 
For small datasets, i.\,e., $131k$ and $262k$ atoms, approach~\ref{en:3} is faster 
than the tree-based approach, since the number of points is too small. For the 
large datasets, the tree approach is faster. In addition, the tree has a smaller 
memory footprint than \texttt{cdist}. This allowed to scale to larger problems, 
e.\,g., a $4M$ atoms and $44.6M$ edges dataset without changing the total number 
of tasks.

Dask shows better scaling than Spark for $131k$, $262k$, and $524k$ atoms. This 
is not true for $4M$ atoms, indicating that Dask's communication layer 
is not able to scale as well as Spark's. Spark shows similar performance with 
MPI4py for the largest dataset due to minimal shuffle traffic. Thus, MPI's efficient 
communication does not become relevant.

\subsection{Conceptual Framework and Discussion}
In this section we provide a conceptual framework that allows application
developers to carefully select a framework according to their requirements
(e.\,g., compute and I/O). It is important to
understand both the properties of the application and Big Data frameworks.
Table~\ref{tab:framework} illustrates the criteria of this conceptual framework
and ranks the three frameworks.
\begin{table}[t]
    \scriptsize
    \centering
    \begin{tabular}{|p{2.9cm}|p{1.5cm}|p{1.3cm}|p{1.3cm}|}
        \hline
        &\textbf{\rp}     &\textbf{Spark} &\textbf{Dask}\\\hline
        \multicolumn{4}{|l|}{\textbf{Task Management}} \\\hline
        Low Latency   &- &o &+\\\hline
        Throughput    &- &+ &++\\\hline
        MPI/HPC Tasks &+ &o &o\\\hline  
        Task API   &+ &o &++\\\hline    
        Large Number of Tasks   &-- &++ &++\\\hline     
        \multicolumn{4}{|l|}{\textbf{Application Characteristics}}\\\hline
        Python/native Code &++ &o &+\\\hline
        
        Java               &o &++ &o\\\hline
        
        Higher-Level Abstraction &- &++ &+\\\hline
        
        Shuffle                  &- &++ &+\\\hline
        
        Broadcast                &- &++ &+\\\hline
        
        Caching                  &- &++ &o\\\hline
    \end{tabular}
    \caption{\textbf{Decision Framework:} Criteria and Ranking for Framework Selection. - : Unsupported or low performance
        +: Supported, ++: Major Support, and o:Minor support.\label{tab:framework}\up}
\end{table}
\upe
\subsubsection{Application Perspective}
We showed that we can implement MD trajectory data analysis applications 
using all three frameworks, as well as MPI4py. Implementation aspects, such as 
computational complexity, and shuffled data size influence the performance greatly. 
For embarrassingly parallel applications with 
coarse grained tasks, such as PSA, the choice of the 
framework does not significantly influence performance (Figures~\ref{fig:HausdorffWrangler} 
and~\ref{fig:comet_wrangler_haus}). In addition, the performance difference against 
MPI4py was not significant (Figures~\ref{fig:HausdorffWrangler} and~\ref{fig:comet_wrangler_haus}
). Thus, aspects, such as programmability and integrate-ability, become more 
important.

For fine-grained data parallelism, a Big Data framework, such as Spark and Dask, 
clearly outperforms \rp (Figures~\ref{fig:All4approachesNoRp},~\ref{fig:rpLF}). 
If coupling is introduced, i.\,e. task communication is required 
(e.\,g., reduce), using Spark becomes advantageous (Approaches~\ref{en:3} 
\& \ref{en:4}). MPI4py outperformed Dask, and Spark, despite both frameworks 
scaling for the larger datasets. Especially Spark was able to provide linear 
speedup for approach~\ref{en:3} of Leaflet Finder (Figure~\ref{fig:All4approachesNoRp}). 
Integrating with frameworks that provide higher level abstractions provides 
scalable solutions for more complex algorithms. However, integrating Spark with 
other tools needs to be carefully considered. The integration of Python tools, 
e.\,g. MDAnalysis, often causes overheads due to the frequent need for serialization 
and copying data between the Python and Java space.

Dask had the smallest learning curve of all three frameworks. As a result, it allows 
for faster prototyping compared to \rp and Spark. \rp's learning curve is more steep, 
but is more versatile than Dask and Spark, by offering the lowest level abstraction. 
Spark had the slowest learning curve. It required tuning to get the number 
of tasks correctly, as well as argument passing to map and reduce functions.\up

\subsubsection{Framework Perspective}
\rp is well suited for HPC applications, e.\,g., ensembles (up to $50k$ 
tasks) of parallel MPI applications, as shown in Ref.~\cite{rp-titan,rp-jsspp18}. 
It has limited scalability when 
supporting large numbers of short-running tasks, as often found in data-intensive 
workloads. The file staging implementation of \rp is not suitable for supporting 
the data exchange patterns, i.e. shuffling, required for these applications. 
However, executing MPI and Spark applications alongside on the same resource 
makes \rp particularly suitable when different programming models need to be 
combined.

Dask provides a highly flexible, low-latency task management and excellent support 
for parallelizing Python libraries. We established that Dask has higher throughput 
(Figures~\ref{fig:dask_spark_rp_wrangler} and~\ref{fig:RP_Dask_Spark_throughput}). 
However, Spark provides better speedups for the largest datasets compared to Dask 
(Figure ~\ref{fig:All4approachesNoRp}). Dask's broadcast (Leaflet Finder approach
~\ref{en:1}) and shuffle (Leaflet Finder approaches~\ref{en:2}-~\ref{en:4}) 
performance is worse for larger problems compared to Spark. Thus, Dask's communication 
layer shows some weaknesses that are particularly visible during broadcast and 
shuffle. Spark needs to be particularly considered for shuffle-intensive applications. 
Its in-memory caching mechanism is particularly suited for iterative algorithms 
that maintain a static set of data in-memory and conduct multiple passes on that 
set.\up
\section{Related Work}
\label{related_work} 
MD analysis algorithms were until recently executed serially and parallelization 
was not straightforward. During the last years several frameworks emerged providing 
parallel algorithms for analyzing MD trajectories. Some of those frameworks are 
HiMach~\cite{himach-2008}, Pteros 2.0~\cite{pteros2015}, MDTraj~\cite{mdtraj-2015}, 
and nMoldyn-3~\cite{nmoldyn-2012}. We compare these frameworks with our approach 
over the parallelization techniques used.

HiMach~\cite{himach-2008} was developed by D. E. Shaw Research group to provide 
a parallel analysis framework for MD simulations, and extends Google's MapReduce. 
HiMach API defines trajectories, does per frame data acquisition (Map) and 
cross-frame analysis (Reduce). HiMach's runtime is responsible to parallelize and 
distribute Map and Reduce phases to resources. Data transfers are done through a 
communication protocol created specifically for HiMach.

Pteros-2.0~\cite{pteros2015} is a open-source library that is used for modeling 
and analyzing MD trajectories, providing a plugin for each supported algorithm. 
The execution is done by a user defined driver application, which setups trajectory 
I/O and frame dispatch for analysis. It offers a C++ and Python API. Pteros 2.0 
parallelizes computational intensive algorithms via OpenMP and Multithreading. As 
a result, it is bounded to execute on a single node, making any analysis execution 
highly dependent on memory size. Through \rp, Spark and Dask, we avoided recompiling 
every time there is a change to the underlying resource, ensuring the application's 
execution.

MDTraj~\cite{mdtraj-2015} is a Python package for analyzing MD trajectories. It 
links MD data and Python statistical and visualization software. MDTraj proposes 
parallelizing the execution by using the parallel package of IPython as a wrapper 
along with an out-of-core trajectory reading method. Our approach allows data 
parallelization on any level of the execution, not only in data read.

nMoldyn-3~\cite{nmoldyn-2012} parallelizes the execution through a Master Worker 
architecture. The master defines analysis tasks, submits them to a task manager, 
which then are executed by the worker process. In addition, it provides adaptability, 
allowing on-the-fly addition of resources, and execution fault tolerance when worker 
processes disconnect.

In contrast, our approach utilizes more general purpose frameworks for parallelization. 
These frameworks provide higher level abstractions, e.g machine learning, so any 
integration with other data analysis methods can be fast and easier. In addition, 
resource acquisition and management is done transparently.
\up
\section{Conclusion and Future Work}
\label{concl}
In this paper, we investigated the use of different programming abstractions and 
frameworks for implementing a range of algorithms for MD trajectory analysis. We 
conducted an in-depth analysis of applications' characteristics and assessed 
the architectures of \rp, Spark and Dask. We provide a conceptual framework that 
enables application developers to qualitatively evaluate task parallel frameworks 
with respect to application requirements. Our benchmarks enable them to 
quantitatively assess framework performance as well as the performance of different 
implementations. Our method can be used for any application which data are represented as 
time series of simulated systems, e.\,g. weather forecast, and earthquakes.

While the task abstractions provided by all frameworks are well-suited for 
implementing all use cases, the high-level MapReduce programming model provided 
by Spark and Dask has several advantages. It is easier to use and efficiently 
support common data exchange patterns, e.\,g. shuffling between \texttt{map} 
and \texttt{reduce} stages. In our benchmarks, Spark outperforms Dask in communication
-intensive tasks, such as broadcasts and shuffles. Further, the in-memory RDD 
abstraction performs well for iterative algorithms. Dask provides more versatile low and 
high level APIs and integrates better with python frameworks. \rp does not provide 
a MapReduce API, but is well suited for coarse-grained task-level parallelism
~\cite{rp-titan,rp-jsspp18}, and when HPC and analytics frameworks need to be 
integrated. We also identified a limitation in Dask and Spark: while both frameworks 
provide some support for linear algebra - both provide a distributed array abstractions 
- it proved inflexible for an efficient all-pairs pattern implementation. 
They required workarounds and utilization of out-of-framework functions 
to read and partition data (Table~\ref{tab:app_operators}). Although, none 
of these frameworks outperformed MPI, their scaling capabilities along with 
their high-level APIs create a strong case on utilizing them for data analytics 
of HPC applications.

In the future, we will further improve the performance of the presented algorithms
, e.\,g., by reducing the memory and computation footprint, data transfer sizes 
between stages, optimizing filesystem usage. To better support PyData tools 
in \rp, we plan to extend the Pilot-Abstraction to support Dask and other Big 
Data frameworks. Thus, providing a system that allows MPI simulations along with 
Big Data frameworks on the same resources. Further, we will refine the \rp task 
execution engine to meet the requirement of data analytics applications and create 
strategies that mitigate issues occurring at large scale, e.\,g. stragglers. 
Another area of research is dynamic resource management and to dynamically scale 
the resource pool (e.\,g., by adding or removing nodes) to meet the requirements 
of a specific application stage.

\footnotesize{{\it Acknowledgements} We thank Andre Merzky
	for useful discussions. This work is funded by NSF 1443054 and 1440677. Computational resources were provided by NSF XRAC awards
	TG-MCB090174 and TG-MCB130177.\footnotesize{{\it Software and Data}Source scripts, data and analysis scripts can be found at: Scripts: \url{http://github.com/radical-cybertools/midas}, Experiments and Data:\url{http://github.com/radical-experiment/midas-exps/}}
\up
\bibliographystyle{ACM-Reference-Format}
 {\scriptsize
\bibliography{local,radical_publications,saga,saga-related,pilotjob,streaming,md,bigdata}}

\end{document}